\documentclass[runningheads]{llncs}
\usepackage{multirow}
\usepackage{booktabs} 
\usepackage{graphicx}
\usepackage[T1]{fontenc}
\usepackage{multirow}
\usepackage{graphicx}
\usepackage{amssymb}
\usepackage{amsmath}
\usepackage{float}
\usepackage{tabularx}
\usepackage{caption}
\usepackage{fixltx2e} 
\graphicspath{ {./images/} }
\usepackage{graphicx}
\usepackage[sorting=none]{biblatex}
\begin{document}
\title{End-to-End User-Defined Keyword Spotting using Shifted Delta Coefficients}
%
%
\author{Kesavaraj V, Anuprabha M, Anil Kumar Vuppala}
\authorrunning{Kesavaraj V et al.}
%


\institute{Speech Processing Laboratory, \\International Institute Of Information Technology Hyderabad, India \\
\email{\{kesavaraj.v, anuprabha.m\}@research.iiit.ac.in}\\
\email{anil.vuppala@iiit.ac.in}}
\maketitle              
\begin{abstract}

Identifying user-defined keywords is crucial for personalizing interactions with smart devices. 
Previous approaches of user-defined keyword spotting (UDKWS) have relied on short-term spectral features such as mel frequency cepstral coefficients (MFCC) to detect the spoken keyword. 
However, these features may face challenges in accurately identifying closely related pronunciation of audio-text pairs, due to their limited capability in capturing the temporal dynamics of the speech signal.
To address this challenge, we propose to use shifted delta coefficients (SDC) which help in capturing pronunciation variability (transition between connecting phonemes) by incorporating long-term temporal information.
The performance of the SDC feature is compared with various baseline features across four different datasets using a cross-attention based end-to-end system. Additionally, various configurations of SDC are explored to find the suitable temporal context for the UDKWS task. The experimental results reveal that the SDC feature outperforms the MFCC baseline feature, exhibiting an improvement of 8.32\% in area under the curve (AUC) and 8.69\% in terms of equal error rate (EER) on the challenging Libriphrase-hard dataset.  
Moreover, the proposed approach demonstrated superior performance when compared to state-of-the-art UDKWS techniques.
\end{abstract}
\section{Introduction}

Advancements in deep learning technology have transformed voice-activated interactions with machines from science fiction to reality.
The proliferation of voice assistants like Amazon's Alexa, Apple's Siri, Google's Assistant, and Microsoft's Cortana are good proof of this \cite{hoy2018alexa}. 
These voice assistants are activated using a technology called spoken keyword spotting, or simply keyword spotting, which detects specific wake-up words within a continuous audio stream \cite{lopez2021deep}. 
This helps to avoid activating the more computationally intensive automatic speech recognition (ASR) when unnecessary.
For instance, Google's voice search responds to the phrase "Okay Google," while Apple's conversational assistant is activated with the phrase "Hey Siri" \cite{vinyals2014chasing}. However, these keywords are not personalized.


With the growing demand for personalized voice assistants, user-defined keyword spotting (UDKWS)  \cite{sacchi2019open,gurugubelli2024comparative}, also known as custom keyword detection or open vocabulary keyword spotting, has gained considerable attention.
Unlike closed vocabulary keyword spotting \cite{sainath15b_interspeech}, where only predetermined keywords are recognized, open vocabulary keyword spotting deals with the challenge of identifying random keywords that the model may not have encountered during training, adding an additional layer of complexity to the task.


 

Over the years, various techniques have been explored for UDKWS. One of the earliest approaches involves the use of large-vocabulary continuous speech recognition (LVCSR) systems \cite{miller2007rapid, chen2013using}. 
These systems decode the speech signal, after which the keyword is searched in the generated lattices.
Another approach is keyword/filler hidden markov model (HMM) \cite{rose1990hidden, rohlicek1989continuous}.
In this approach, separate HMMs are trained to model keyword and non-keyword audio segments.  
While these architectures allow for customization of the keyword by modifying the decoding graph, the computational requirements remain significant.



Recent works in UDKWS have focused on developing end-to-end systems that take two inputs: the enrolled keyword references and the speech data to be detected. One such classical approach is the query-by-example (QbyE)~\cite{huang2021query, lugosch2018donut} approach, which involves matching input queries with pre-enrolled examples. However, the effectiveness of the QbyE method relies heavily on the similarity between the recorded speech during enrollment and the subsequent evaluated speech recordings.  Challenges such as diverse vocal characteristics among users and background noise in different environments can significantly impact the consistency of the QbyE method's performance. To address these challenges, researchers have explored text enrollment-based methods \cite{sacchi2019open, shin22_interspeech}. 

For instance, \cite{audhkhasi2017end} proposes an ASR-free end-to-end system that generates audio embedding and keyword embedding using an acoustic encoder and a keyword encoder, respectively. These embeddings are then merged into a multilayer perceptron for keyword existence prediction.
In \cite{shin22_interspeech}, an attention-based cross-modal matching approach is proposed to learn the agreement between audio and text
keyword at the utterance level. In \cite{lee23d_interspeech}, a novel zero-shot UDKWS is proposed to learn the audio-phoneme relationship of the keyword through phoneme-level detection loss. Also, \cite{nishu23_interspeech} introduced dynamic sequence partitioning to optimally partition the audio embedding sequence into the same length as the text sequence. These recent end-to-end techniques \cite{lee23d_interspeech, nishu23_interspeech, shin22_interspeech}, primarily depend on evaluating speech and text representations in a common latent space, demonstrated promising results in the custom keyword spotting task. 
Despite these significant advancements in the field of UDKWS, the predominant focus has been on the advancement of deep learning models and training approaches. There has been relatively limited exploration of feature engineering which plays a crucial role in enhancing the performance of any speech application.
This observation has motivated us to perform feature-level exploration for UDKWS task.

In literature, mel-scale related features such as MFCC and mel spectrogram are the most commonly used features in UDKWS \cite{shin22_interspeech, nishu23_interspeech, lee23d_interspeech}. 
While these features provide a good estimation of the local spectra, they may not fully capture the temporal dynamics, such as changes in pronunciation over time, present in spoken keywords.
Incorporating contextual information could help the system in capturing this pronunciation variability by modeling frame-level dependencies. 
Notably, in some studies \cite{kumarvuddagiri18_sltu, torres2002approaches}, SDC is well-regarded for their ability to capture long-term temporal information (stacking delta features across several frames) for language identification tasks.
Motivated by this fact, we propose to use the SDC feature to enhance the robustness of UDKWS, especially in distinguishing between similar pronunciations of audio-text pairs.
To the best of our knowledge, this is the first study that explored the importance of long-term temporal information at feature level for UDKWS task.
The key contributions of this study include:
\begin{itemize}
\item Performance comparison of SDC features with commonly used short-term spectral features, namely MFCC, mel spectrogram, perceptual linear prediction coefficients (PLP), and relative spectral-perceptual linear prediction coefficients (RASTA-PLP) in a common experimental setup.
\item Exploration of different configurations of SDC features to determine the appropriate temporal context for the UDKWS task.


\item Examination of the system’s performance for keywords of different word lengths.
\item Assessment of the efficiency of the proposed approach through comprehensive comparisons with various state-of-the-art UDKWS systems.

\end{itemize}

The organization of this paper is as follows: Section~\ref{sec:Architecture} provides details about the architecture, Section~\ref{sec:Feature}  discusses the feature extraction techniques, Section~\ref{sec:Experimental Setup} describes the experimental setup, Section~\ref{sec:Results} presents the results and discussion and Section~\ref{sec:Conclusion} concludes the study.

\section{Architecture}
\label{sec:Architecture}
\begin{figure*}[!htbp]
  \centering
  \includegraphics[width=\textwidth]{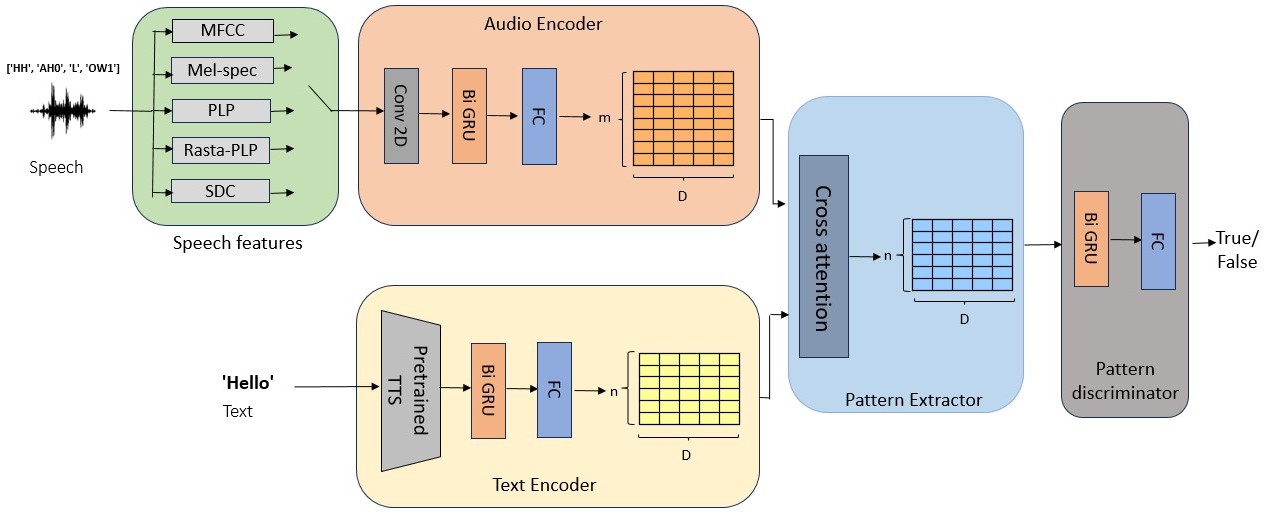}
  \caption{Proposed architecture for user-defined keyword spotting}
  \label{fig:architecture}
\end{figure*}

In this section, we will discuss the details of the architecture that is used for studying different audio features for the UDKWS task, as shown in Fig.~\ref{fig:architecture}. The proposed architecture is adopted from \cite{v2024open}. It consists of four submodules: audio encoder, text encoder, pattern extractor, and pattern discriminator. 

\subsection{Audio Encoder}
In this study, various short-term and long-term spectral features (discussed in Section~\ref{sec:Feature}) are used as input to the audio encoder. The encoder utilized two 2-D convolutional layers (Conv2D), each consisting of 32 filters with a kernel size of 3. To enhance computational efficiency, the initial convolution layer employs a stride of 2 to skip processing of  consecutive frames. Additionally, batch normalization is applied after each Conv2D operation to ensure stable training.  Following the Conv2D layers, two bidirectional gated recurrent units (Bi-GRU) with a dimension of 64 each are utilized. Finally, a 128-dimensional audio embedding is produced by passing the output from the final Bi-GRU layer to a dense layer. The output from the audio encoder is denoted as \(E_a \in \mathbb{R}^{m\times D}\), where m and D denote the length of the audio (i.e. number of frames) and 
embedding dimension, respectively.
 

\subsection{Text Encoder}
It includes a pre-trained Tacotron 2 \cite{shen2018natural} model, a recurrent sequence-to-sequence text-to-speech (TTS) system, which takes character sequences as input to produce the corresponding audio output.
The integration of a pre-trained TTS system in the text encoder is inspired by \cite{v2024open}. The main motivation for including the TTS system is to generate text representations that are aware of audio projections.
According to \cite{v2024open}, employing intermediate representations from the pre-trained TTS model leads to superior performance compared to using character embeddings as text features. It is also demonstrated that representations from the LSTM block of the Tacotron 2 encoder, with a dimension of 512, effectively act as text features. Consequently, this study adopts a similar approach. The resulting intermediate representations from the TTS model are then passed through a Bi-GRU layer with a dimension of 64. The output from the Bi-GRU layer is subsequently fed into a dense layer with 128 units. The output from the text encoder is denoted as \(E_t \in \mathbb{R}^{n \times D}\), where n and D denote the length of the text (i.e. number of characters) and embedding dimension, respectively.


\subsection{Pattern Extractor}
Motivated by \cite{vaswani2017attention}, the pattern extractor employs a cross-attention mechanism to capture temporal correlations between audio and text embeddings.
All hidden states from the output of audio and text encoders are fed into the cross-attention layer to preserve the temporal information.
In this setup, the audio embedding $E_a$ functions as both the key and value, while the text embedding $E_t$ acts as the query. 
The resulting context vector from the pattern extractor contains information regarding the agreement between audio and text. 

\subsection{Pattern Discriminator}
The pattern discriminator determines whether audio and text inputs share the same keyword or not. 
To achieve this, it consists of a single Bi-GRU layer with a dimension of 128 that takes the context vector from the pattern extractor as input. Subsequently, the output from the last frame of the Bi-GRU layer is passed through a dense layer with a sigmoid as an activation function.

\section{Feature Extraction}
\label{sec:Feature}
Representing a speech utterance in a vector of parameters is defined as feature extraction \cite{deshwal2020language}. The significant aim of performing this step is to derive the appropriate/relevant information.
In this section, we discuss about five important feature extraction techniques, namely mel spectrograms \cite{shin22_interspeech}, MFCC \cite{abdul2022mel}, PLP \cite{hermansky1990perceptual}, RASTA-PLP \cite{hermansky1992rasta}, SDC \cite{kumarvuddagiri18_sltu}. These five features are selected after reviewing several works \cite{meghanani2021exploration, palo2015use, milner2002comparison, upadhya2018thomson, sahidullah15_interspeech} on speech related applications.

\subsection{Mel Spectrogram}
The mel spectrogram is derived from the magnitude spectrogram, but it's different because the mel filter bank mimics the human ear's perception and emphasizes the lower frequency region more than the higher frequencies. 
Initially, the magnitude spectrogram—time-frequency representation—is obtained by segmenting the speech waveform into windowed segments and applying the fast fourier transform to each segment.
Following this, the computed magnitude spectrogram is mapped to a mel-scale using 40 mel filter banks, and then subjected to a logarithmic operation to generate the mel-spectrogram.

\subsection{Mel-Frequency Cepstral Coefficients} 
MFCC, a popular feature in speech signal processing, captures vocal tract characteristics by representing the short-term power spectrum through a linear cosine transform. This transformation operates on a logarithmic power spectrum, nonlinearly scaled to the mel frequency range. The process of calculating MFCC involves windowing the signal, calculating discrete Fourier transform (DFT) coefficients for each window, taking the logarithm of the DFT magnitude, filtering frequencies with the mel scale, and finally extracting the MFCC coefficients.
The first 13 cepstral coefficients are considered as MFCC features. 
The first and second derivatives of 13-dimensional MFCC features have been combined with static MFCCs, referred to as MFCC+$\Delta$+$\Delta\Delta$, to capture the temporal dynamics present in the speech signal.


\subsection{Perceptual Linear Prediction}

Several alternatives to MFCC have been proposed for representing short-term speech signals - One such alternative is PLP (perceptual linear prediction coefficients). 
PLP is a feature that gives representation conforming to a smoothed short-term spectrum that has been equalized and compressed, similar to human hearing, thus making it similar to the MFCC. 
The process of calculating PLP features starts with windowing the signal, computing discrete Fourier transform (DFT) coefficients for each window, and taking the logarithm of the DFT magnitude to obtain power spectral estimates. 
Next, a trapezoidal filter is applied at 1-bark intervals to merge overlapping critical band filter responses in the power spectrum, thereby compressing higher frequencies into a narrow band.
Finally, the spectral amplitude is compressed by taking the cubic root to match the nonlinear relationship between sound intensity and perceived loudness.

\subsection{Relative Spectral -  Perceptual Linear Prediction}

RASTA-PLP builds upon the PLP technique by introducing a crucial addition: a bandpass filter at each sub-band. By suppressing undesirable frequencies, RASTA-PLP increases the robustness of PLP to noise. The process of RASTA-PLP involves several steps. Initially, it calculates the critical-band power spectrum, followed by the application of a compressing static nonlinear transformation to the spectral amplitude. Subsequently, it filters the time trajectory of each transformed spectral component with a bandpass filter. After this, the filtered speech undergoes further transformation using an expanding static nonlinear transformation. Additionally, it includes equal loudness curve adjustment and the application of the intensity-loudness power law to replicate the human auditory system. In essence, RASTA filtering acts as a modulation-frequency bandpass filter, emphasizing the modulation frequency range most relevant to speech, while disregarding lower or higher modulation frequencies. 



\subsection{Shifted Delta Coefficients}
SDC features, extensively utilized in language identification \cite{torres2002approaches}, are pivotal for capturing long-term temporal information. Inspired by this, we propose to use the SDC features for the UDKWS task. The main motivation is to enhance the model's ability to capture the pronunciation variability of the spoken keyword, thus improving the overall performance of UDKWS.
In this study, SDC features are computed from Mel-spectrogram, since it gives better performance compared to all other short-term spectral features. The calculation of the SDC feature depends on four parameters, and it is represented as N-d-p-k. 
Here N represents the number of cepstral coefficients for every frame, d denotes the amount of shift (delay) from the current frame, p denotes the shift between the consecutive delta blocks, and k denotes the number of frames whose deltas are to be concatenated. The delta feature vector for \(t^{th}\) frame in the \(i^{th}\) iteration is computed as

\begin{equation}
\delta_c(t, i) = c(t + ip + d) - c(t + ip - d), \quad \text{where } 0 \leq i \leq k-1
\end{equation}

These k delta computations are stacked as in (\ref{tab:stacked SDC})
to form k×N dimensional shifted delta coefficients

\begin{equation}
\label{tab:stacked SDC}
SDC(t) = 
\begin{pmatrix}
\delta_c(t, 0) \\
\delta_c(t, 1) \\
\vdots \\
\delta_c(t, k - 1)
\end{pmatrix}
\end{equation}

The computed stacked delta features are then combined with the static mel spectrogram features to obtain the final SDC feature vector. 
The effect of variation in parameters (d and k), which control the amount of temporal context, has also been studied.

\section{Experimental Setup}
\label{sec:Experimental Setup}
\subsection{Database}
The LibriPhrase dataset \cite{shin22_interspeech}, derived from the LibriSpeech corpus \cite{panayotov2015librispeech}, is utilized for both training and evaluation. It comprises short phrases with varying word lengths, ranging from 1 to 4.
The training set of LibriPhrase was generated using the train-clean-100 and train-clean-360 subsets, while the evaluation set was derived from the train-others-500 subset. The evaluation set consists of 4391, 2605, 467, and 56 episodes of each word length respectively. Each episode has three positive and three negative pairs. The negative samples are further categorized into easy and hard based on Levenshtein distance \cite{levenshtein1966binary}, leading to the creation of the LibriPhrase Easy (LP\textsubscript{E}) and LibriPhrase Hard (LP\textsubscript{H}) datasets.  
Each example is denoted by 3 entities: (audio, text, target) where the target value is 1 for a positive pair and 0 for a negative pair.

For a comprehensive evaluation of model performance, we expanded our assessment beyond the LibriPhrase dataset by including two additional datasets: the Google Speech Commands V1 dataset (G) \cite{warden2018speech} and the Qualcomm Keyword Speech dataset (Q) \cite{kim2019query}. The Google Speech Commands V1 dataset (G) contains speech recordings from 1,881 speakers, emphasizing 30 small keywords. From that, the validation dataset corresponding to 30 keywords is used for evaluation. On the other hand, the Qualcomm Keyword Speech dataset (Q) includes 4,270 utterances of four keywords spoken by 50 speakers. Each speaker in this dataset contributes approximately 22-23 instances for each keyword.

\subsection{Implementation Details}
In the feature extraction, all spectral vectors are obtained by block processing the whole speech into short segments using a window length of 25 ms and overlap of 10 ms. A pre-emphasis factor of 0.97 is applied to boost the amount of energy in the high frequencies. Hamming window is used during the windowing process of speech signal to reduce the spectral leakage. Zero padding is applied
along the time dimension to ensure that the input feature representation is of equal size, as required by the input of the Conv2D layer.

The training pipeline is structured as a binary classification task with the objective of classifying the similarity of input pairs \{audio, text\}.
A dropout of 0.2 is applied after each layer in both audio and text encoders to prevent overfitting. 
The training process utilizes binary cross-entropy loss as the training criterion and employs the Adam optimizer \cite{kingma2014adam} with default parameters for optimization.
A batch size of 128 and a fixed learning rate of $10^{-4}$ are used for training the model and, 
the best model is selected based on its performance on the validation set. For training, we used four NVIDIA GeForce RTX 2080 Ti GPUs.

\section{Results and discussions}
\label{sec:Results}
\vspace{-0.05 cm}
The proposed approach provides comprehensive insights about the spoken keyword by leveraging SDC features, known for their proficiency in capturing long-term temporal information. 
To validate that we conducted extensive experiments across diverse datasets, with results and plots presented in this section.

\vspace{-0.1 cm}
\subsection{Comparison of different front-end features}
In this section, the performance comparison between SDC features and various short-term spectral features is discussed to study the importance of long-term temporal information in the UDKWS task. The results are presented in Table~\ref{tab:comparison of features}.
Upon analysing the results, it's clear that SDC features consistently outperform all baseline features across all datasets.
Furthermore, when compared to the MFCC feature, the SDC feature demonstrates a notable improvement of 8.69\% in AUC and 8.32\% in EER on the challenging LP\textsubscript{H} dataset, which comprises similar pronunciations of audio-text pairs (e.g., "madame" and "modem").
This improvement is attributed to their ability to capture the temporal dynamics of speech signals by incorporating contextual information. 
On the other hand, PLP, RASTA-PLP, and mel spectrogram exhibit competitive performance with consistently good AUC scores, indicating them as preferable alternatives to SDC. In contrast, MFCC performs poorly compared to all other features. Compared to MFCC alone, MFCC+$\Delta$+$\Delta\Delta$ which models the temporal dynamics of the speech signal to some extent by concatenating their first and second derivatives to the original MFCC feature, provides significant improvements. Overall, the observations from Table~\ref{tab:comparison of features} suggest that  speech information extracted over a longer
context 
plays a pivotal role in the development of systems for UDKWS tasks.
\renewcommand{\arraystretch}{1.2}

\begin{table*}[!ht]
\caption{Performance comparison of  SDC features with various short-term spectral features across different datasets: Google Commands V1 (G), Qualcomm Keyword Speech dataset (Q), Libriphrase-Easy (LP\textsubscript{E}), and Libriphrase-Hard (LP\textsubscript{H})}
\vspace{0.2 cm}
\setlength{\tabcolsep}{4.5 pt}
\label{tab:comparison of features}
\small
\centering
\begin{tabular}{|c|cccc|cccc|}
\hline
\multirow{2}{*}{\textbf{Features}} & \multicolumn{4}{c|}{\textbf{EER (\%)}}                                                                                          & \multicolumn{4}{c|}{\textbf{AUC (\%)}}                                                                                           \\ \cline{2-9} 
                                   & \multicolumn{1}{c|}{\textbf{G}}     & \multicolumn{1}{c|}{\textbf{Q}}     & \multicolumn{1}{c|}{\textbf{LP\textsubscript{E}}}  & \textbf{LP\textsubscript{H}}   & \multicolumn{1}{c|}{\textbf{G}}     & \multicolumn{1}{c|}{\textbf{Q}}     & \multicolumn{1}{c|}{\textbf{LP\textsubscript{E}}}   & \textbf{LP\textsubscript{H}}   \\ \hline
MFCC                               & \multicolumn{1}{c|}{32.24}          & \multicolumn{1}{c|}{12.59}          & \multicolumn{1}{c|}{7.99}          & 29.8           & \multicolumn{1}{c|}{73.95}          & \multicolumn{1}{c|}{91.2}           & \multicolumn{1}{c|}{97.8}           & 77.21          \\ \hline
MFCC+$\Delta$+$\Delta\Delta$
                  & \multicolumn{1}{c|}{30.28}          & \multicolumn{1}{c|}{11.8}           & \multicolumn{1}{c|}{6.8}           & 27.01          & \multicolumn{1}{c|}{76.5}           & \multicolumn{1}{c|}{93.29}          & \multicolumn{1}{c|}{98.06}          & 78.54          \\ \hline
Mel Spectrogram                    & \multicolumn{1}{c|}{27.91}          & \multicolumn{1}{c|}{16.7}           & \multicolumn{1}{c|}{5.89}          & 26.45          & \multicolumn{1}{c|}{79.13}          & \multicolumn{1}{c|}{90.97}          & \multicolumn{1}{c|}{98.21}          & 79.81          \\ \hline
PLP                                & \multicolumn{1}{c|}{28.43}          & \multicolumn{1}{c|}{15.37}          & \multicolumn{1}{c|}{6.58}          & 25.22          & \multicolumn{1}{c|}{77.65}          & \multicolumn{1}{c|}{90.47}          & \multicolumn{1}{c|}{97.88}          & 78.81          \\ \hline
RASTA-PLP                          & \multicolumn{1}{c|}{27.4}           & \multicolumn{1}{c|}{14.32}          & \multicolumn{1}{c|}{6.42}          & 25.84          & \multicolumn{1}{c|}{78.05}          & \multicolumn{1}{c|}{91.24}          & \multicolumn{1}{c|}{97.82}          & 79.7           \\ \hline
SDC                                & \multicolumn{1}{c|}{\textbf{23.54}} & \multicolumn{1}{c|}{\textbf{9.61}} & \multicolumn{1}{c|}{\textbf{3.84}} & \textbf{21.48} & \multicolumn{1}{c|}{\textbf{83.56}} & \multicolumn{1}{c|}{\textbf{96.73}} & \multicolumn{1}{c|}{\textbf{98.34}} & \textbf{85.90} \\ \hline
\end{tabular}%
\end{table*}

\vspace{0.2 cm}
\subsection{SDC configuration}

\begin{figure}[]
  \centering
  \vspace{-0.4 cm}
  \includegraphics[width=0.9\linewidth]{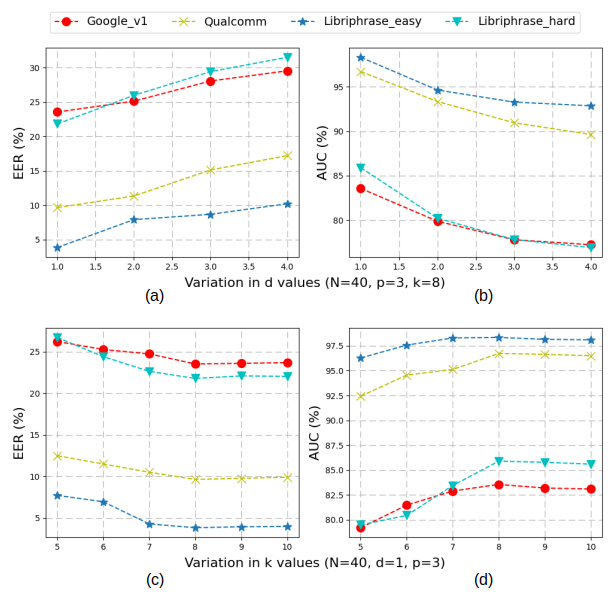}
  \caption{Performance Analysis of SDC Configuration across four datasets. (a) \& (b) illustrate the effect of varying d values. (c) \& (d) illustrate the effect of varying k values.}
  \label{fig: plot}
\end{figure}

The calculation of SDC features relies on four parameters: N, d, p, and k. 
Variations in these parameters can significantly affect the amount of temporal context captured.
Therefore, an ablation study is conducted by varying d and k values to determine the suitable temporal context for UDKWS, with the results depicted in Fig.~\ref{fig: plot}. Fig.~\ref{fig: plot} (a) \& (b) illustrates the performance change in terms of AUC and EER for varying d values (amount of shift from the current frame) from 1 to 4 while keeping the other parameters fixed. It can be observed that as the d value increases, the performance drops across all datasets, indicating a loss of information when the shift is increased in calculating SDC features. Fig.~\ref{fig: plot} (c) \& (d) show the effect of varying k values (number of frames whose deltas are stacked) ranging from 5 to 10, while keeping the other parameters fixed. It is evident that as the k value increases from 5 to 8, the performance in terms of AUC and EER also increases across all datasets, demonstrating the effect of the context window on UDKWS. 
However, beyond a k value of 8, performance either saturates or declines, indicating the need for precise contextual information to reliably recognize keywords.
Overall, it is evident that the SDC configuration 40-1-3-8 exhibits the best performance with optimal temporal context for UDKWS.

\vspace{-0.2 cm}
\subsection{Analysis on the length of keywords}
\vspace{-0.5 cm}
\begin{table*}[htbp]
\caption{Performance comparison of SDC and mel spectrogram features across different word lengths}
\vspace{0.2 cm}
\setlength{\tabcolsep}{4 pt}
\label{tab:word length}
\small
\centering
\begin{tabular}{|c|c|c|c|c|}
\hline
\textbf{Feature}                 & \textbf{Word Length} & \textbf{EER (\%)} & \textbf{AUC (\%)} & \textbf{F 1 score (\%)} \\ \hline \hline
\multirow{4}{*}{Mel Spectrogram} & 1                    & 7.67              & 97.01             & 91.11          \\ \cline{2-5} 
 & 2 & 8.55 & 96.57 & 90.98 \\ \cline{2-5} 
 & 3 & 9.05 & 95.6 & 89.76 \\ \cline{2-5} 
 & 4 & 9.25 & 95.34  & 88.30 \\ \hline \hline
\multirow{4}{*}{SDC}             & 1                    & 5.37              & 98.25             & 94.34          \\ \cline{2-5} 
 & 2 & 6.35 & 97.87 & 93.31 \\ \cline{2-5} 
 & 3 & 7.24 & 96.91 & 91.57 \\ \cline{2-5} 
 & 4 & 8.29 & 96.31 & 90.32 \\ \hline
\end{tabular}%
\end{table*}

The evaluation of the system across different word lengths for mel spectrogram (best performing among baseline) and SDC features is studied on the LibriPhrase evaluation dataset. The results in terms of EER, AUC and F1-score are presented in Table~\ref{tab:word length}. 
When compared to the mel spectrogram, SDC demonstrates absolute improvements of 3.24\%, 2.33\%, 1.81\%, and 2.02\% in F1-score for word lengths 1, 2, 3, and 4 respectively.
Despite the improvements, SDC encounters challenges similar to the mel spectrogram in keyword recognition as word length increases. 

\vspace{1 cm}
\subsection{Comparison of various UDKWS techniques}
\vspace{-0.25 cm}
In this section, the performance of the proposed approach is compared with various state-of-the-art UDKWS techniques across G, Q, LP{\textsubscript{E}}, {LP{\textsubscript{H}}} datasets, and the results are presented in Table~\ref{tab:baseline comparison}.
Evaluation results show that among all the baselines, CMCD (cross-modlaity correspondence detector) \cite{shin22_interspeech}
demonstrates strong performance, while Triplet \cite{sacchi2019open} shows weak
performance across the Q, LP\textsubscript{E}, and LP\textsubscript{H} datasets.
The attention-based QbyE method demonstrates superior performance on the G dataset due to its similarity scoring mechanism, particularly when the keyword is included in the training set.
However, it
shows degraded performance when the keyword is unfamiliar as observed in Q and LibriPhrase. 
In contrast, our proposed approach outperforms all the baselines on all datasets except G. 
Specifically, compared to the CMCD baseline, our proposed method demonstrates a substantial improvement in EER of 4.58\% in the LP\textsubscript{E} dataset and 11.42\% in the LP\textsubscript{H} dataset.
Additionally, the model is evaluated on datasets G and Q without any fine-tuning to assess its generalization capability.
\begin{table*}[!ht]
\vspace{-0.9 cm}
\caption{Performance comparison of the proposed method with various UDKWS techniques across different datasets: Google Commands V1 (G), Qualcomm Keyword Speech dataset (Q), LibriPhrase-Easy (LP\textsubscript{E}), and LibriPhrase-Hard (LP\textsubscript{H}).
}
\vspace{0.2 cm}
\setlength{\tabcolsep}{4.5 pt}
\label{tab:baseline comparison}
\small
\centering
\begin{tabular}{|c|cccc|cccc|}
\hline
\multirow{2}{*}{\textbf{Method}} & \multicolumn{4}{c|}{\textbf{EER (\%)}}                                                                                      & \multicolumn{4}{c|}{\textbf{AUC (\%)}}                                                                                       \\ \cline{2-9} 
                                 & \multicolumn{1}{c|}{\textbf{G}} & \multicolumn{1}{c|}{\textbf{Q}}     & \multicolumn{1}{c|}{\textbf{LP\textsubscript{E}}}  & \textbf{LP\textsubscript{H}}   & \multicolumn{1}{c|}{\textbf{G}} & \multicolumn{1}{c|}{\textbf{Q}}     & \multicolumn{1}{c|}{\textbf{LP\textsubscript{E}}}   & \textbf{LP\textsubscript{H}}   \\ \hline
CTC \cite{lugosch2018donut}                              & \multicolumn{1}{c|}{31.65}      & \multicolumn{1}{c|}{18.23}          & \multicolumn{1}{c|}{14.67}         & 35.22          & \multicolumn{1}{c|}{66.36}      & \multicolumn{1}{c|}{89.69}          & \multicolumn{1}{c|}{92.29}          & 69.58          \\ \hline
Attention \cite{huang2021query}                        & \multicolumn{1}{c|}{\textbf{14.75}}      & \multicolumn{1}{c|}{49.13}          & \multicolumn{1}{c|}{28.74}         & 41.95          & \multicolumn{1}{c|}{\textbf{92.09}}      & \multicolumn{1}{c|}{50.13}          & \multicolumn{1}{c|}{78.74}          & 62.65          \\ \hline
Triplet \cite{sacchi2019open}                         & \multicolumn{1}{c|}{35.6}       & \multicolumn{1}{c|}{38.72}          & \multicolumn{1}{c|}{32.75}         & 44.36          & \multicolumn{1}{c|}{71.48}      & \multicolumn{1}{c|}{66.44}          & \multicolumn{1}{c|}{63.53}          & 54.88          \\ \hline
CMCD  \cite{shin22_interspeech}                           & \multicolumn{1}{c|}{27.25}      & \multicolumn{1}{c|}{12.15}          & \multicolumn{1}{c|}{8.42}          & 32.9           & \multicolumn{1}{c|}{81.06}      & \multicolumn{1}{c|}{94.51}          & \multicolumn{1}{c|}{96.7}           & 73.58          \\ \hline
Proposed                         & \multicolumn{1}{c|}{23.54}      & \multicolumn{1}{c|}{\textbf{9.61}} & \multicolumn{1}{c|}{\textbf{3.84}} & \textbf{21.48} & \multicolumn{1}{c|}{83.56}      & \multicolumn{1}{c|}{\textbf{96.73}} & \multicolumn{1}{c|}{\textbf{98.34}} & \textbf{85.9} \\ \hline
\end{tabular}%
\end{table*}
We observe a consistent improvement of approximately 2\% on the AUC metric and 3\% on the EER metric across datasets G and Q compared to the CMCD baseline. This demonstrates the effectiveness of the proposed approach in recognizing user-defined keywords that are not seen during training.

\vspace{-1 cm}
\section{Conclusion}
\vspace{-0.2 cm}
\label{sec:Conclusion}
This study presented the importance of long-term temporal information for the UDKWS task.
The evaluation results indicated that SDC features outperformed the widely used short-term spectral features. Notably, it showcased its potential in distinguishing similar pronunciations of audio-text pairs in the Libriphrase hard dataset. Furthermore, the ablation study on different SDC configurations revealed that configuration 40-1-3-8 exhibited the best performance with a suitable temporal context. Moreover, the proposed approach demonstrated superior performance compared to state-of-the-art UDKWS approaches. 
In future work, the focus will be on improving the performance of the UDKWS system by exploring the potential of hybrid feature extraction approaches rather than relying on individual counterparts.
\printbibliography

\end{document}